# TYPE VARIABILITY AND COMPLETENESS OF INTERFACES IN JAVA APPLICATIONS

Hani Abdeen and Osama Shata

Department of Computer Science Engineering, Qatar University, Doha, Qatar

*ABSTRACT*

*Interfaces are widely used as central design elements of Java applications. Although interfaces are abstract types similar to abstract classes, the usage of interfaces in Java applications may considerably differ from the usage of abstract classes. Unlike abstract classes, interfaces are meant to enable multiple inheritance in Java programs. Hence, interfaces are meant to encode shared similarities between classes belonging to different class-type hierarchies. Therefore, it is frequent to use interfaces as partial types, where an interface specifies one specific aspect or usage of its implementing classes. In this paper, we investigate interfaces' usage in Java applications from two perspectives. First, we investigate the usage of interfaces as types of classes belonging to different class-type hierarchies (i.e., interface's type variability). Second, we investigate the usage of interfaces as partial types of implementing classes (i.e., interface's type completeness).*

## 1. INTRODUCTION

Software Interfaces represent abstract service contracts governing interactions between logic modules. Formally, they are reference types used to encode similarities among classes of different types. Interfaces are widely used as central design elements of Java applications. Although interfaces are abstract types similar to pure abstract classes, the usage of interfaces in Java applications may considerably differ from the usage of abstract (pure abstract) classes [1], [2]. Unlike classes, and abstract classes, interfaces are meant to enable multiple inheritance in Java programs. Hence, interfaces are meant to encode shared similarities between classes belonging to different class-type hierarchies. Therefore, it is frequent to use interfaces as partial types, where an interface specifies one specific aspect or usage of its implementing classes [3] [4]. A good design of interfaces is a fundamental key to understand the whole application services, and the interactions between modules and subsystems [8]. Despite the importance of interfaces, only a few research efforts investigate interfaces' design in Java applications. Most existing work largely investigated program design at the class level without focusing on the specifics of interfaces [8].

**Contributions**. In this paper, we investigate interfaces' usage in Java applications from two perspectives:
- We investigate the usage of interfaces as types of classes belonging to different classtype   hierarchies (i.e., interface's type variability).
- We investigate the usage of interfaces as partial types of implementing classes (i.e., interface's type completeness).

DOI : 10.5121/ijsea.2014.5301            1

The results show that interfaces are, surprisingly, often used as types for classes belonging to the same class-type. Indeed, a considerable body of analyzed interfaces are implemented by only one implementing class. Hence, their type variability is null. In the same vein, the results show that interfaces are most frequently used as *complete* types of their implementing classes.

## 2. VOCABULARIES

Java tutorial defines an interface as a reference type that, in its most common form, declares a group of related methods with empty bodies. An interface can be implemented by different classes that do not, necessarily, belong to the same hierarchy. A class may implement different interfaces. Indeed, interfaces enable multi-inheritance mechanism in Java programs. The type variability of an interface is hence defined by the variability of types (i.e., classes) which implement that interface (see Section 3).

It is frequent to see classes that are used by their clients for accessing specific functionality of those used classes [4]. Hence, it is better to restrict the access to those functionality that are actually required by the clients. For this purpose, the type of variables (or access points) should be declared as interfaces which specifies only the functionality that is actually used in the contexts clients. Such interfaces are usually defined as partial types since they focus on a specific functionality of their implementing classes.

## 3. INTERFACE'S TYPE VARIABILITY

Since interfaces are meant to enable multi-inheritance in Java programs, they are expected to be used mainly to encode similarities among classes of different types. To measure this expectation of interfaces we define the Type Variability measurement, *TV*. Let *IC(i)* denote the set of all implementing classes of the interface *i*. Let *root_type(c)* be the predicate which returns the root class-type of the class *c*. The type variability of an interface *i*, *TV(i)*, is defined by the number of *different* root-types of its implementing classes:

$$TV(i) = |\bigcup_{c_i \in IC(i)} root\_type(c_i)|$$

According to the above-mentioned definition, if an interface *i* is implemented by 3 different classes (*c1*, *c2*, and *c3*), and if *c1* and *c2* belong to the same class-type hierarchy (i.e., inherit from the same root-type *t1* --which could be any class except the default root class in Java, *Object*), but the class c3 belong to another root-type, *t2*, then the *TV* of *i* is equal to 2. It is worth noting that classes which do not inherit from other classes (i.e., they are direct subclasses of *Object*) are by default different root-types. The root-types of such classes are the classes themselves.

An interface that all its implementing classes belong to (inherit from) the same root-type has thus the minimal type variability, which is 1 --this is by excluding interfaces which are not implemented, at all, and which their *TV* value is 0.

The larger the *TV* value for an interface is, the larger the usage context of the interface. An interface that has a *TV* value higher than 1 indicate a place where the usage of interface concept instead of (abstract) class concept is well justified to enable multi-inheritance. Interfaces which have a *TV* value equal to 1 indicate places where the interface concept might be overused: where using the abstract-class concept might be more appropriate.

## 4. INTERFACE'S TYPE COMPLETENESS

As discussed earlier, interfaces are also meant to restrict the access of clients to specific functionality of used classes, the type of access points should be declared as interfaces which specifies only the functionality that is actually used in the contexts clients. Such interfaces are usually defined as partial types since they focus on a specific functionality of their implementing classes. To measure the type partiality/completeness of an interface, we define the Type Completeness measurement, *TC*. Let *PM(x)* denote the set of all public methods declared/implemented in the interface/class *x*. Let *IC(i)* denote the set of all implementing classes of the interface *i*. The type completeness of an interface *i* is defined as the average ratio of number of declared methods in *i* to the number of public implemented methods in its implementing classes:

$$TC(i) = \frac{\sum_{c_i \in IC(i)} \frac{|PM(i)|}{|PM(c_i)|}}{|IC(i)|}$$

According to the above definition, if an interface i declares 2 methods is implemented by 2 classes, c1 and c2, that each one implement the interface methods, and defines additional methods as follows: c1 defines in total 2 public methods (only the methods that are declared in *i*); c2 defines in total 4 public methods (2 methods in addition to the methods that are declared in *i*). The *TC* value for *i* is thus: *((2/2) + (2/4)) / 2 = 0.75*. This shows that *i* can be considered somewhat as a complete type rather than as a partial type. Indeed, *i* is implemented by only 2 classes and it declares all the public methods of its first implementing class, but only 50% of the public methods of its second implementing class.

The *TC* metric takes its values in [0..1], where the larger the value of *TC* is, the higher the type completeness of the considered interface. An interface with a *TC* value that is equal to 1 is thus said a complete type interface since it declares all the functionality of its implemented classes. Such an interface is expected to violate the Role Decoupling pattern [4].

## 5. STUDY

Our study aims at investigating the design and use of interfaces in Java applications from two perspectives:

- Interface's Type Variability: we investigate whether interfaces are used as types of classes belonging to different class-type hierarchies (to enable multi-inheritance), or they are overused to encode shared similarity between classes belonging to the same class family.

- Interface's Type Completeness: we investigate the usage of interfaces as partial types of implementing classes.

## 5.1. Case-studies

We conduct our study analysing interfaces design an usage in three well-known Java opensource applications: ArgoUML is a UML modelling tool; Hibernate is a Relational persistence for idiomatic Java; JHotDraw is a graphics framework for structured drawing editors. Table 1 shows additional information about the number of interfaces, classes and implemented classes considered in each case-study.

Table 1. Information about studied software projects.

| Application | Interfaces | Classes | Implementations |
|---|---|---|---|
| ArgoUML | 88 | 1793 | 358 |
| Hibernate | 319 | 2658 | 745 |
| JHotdraw | 42 | 396 | 169 |

## 5.2. Results Analysis

- *Interface Type Variability:* Figure 1 shows the histogram of interface type variability in analyzed applications: (null, TV = 1) (tiny, TV = 2); (small, $2 < TV \leq 5$); (medium, $5 < TV \leq 10$); (large, $10 < TV \leq 15$); (huge, $15 < TV$). The figure shows that for all analyzed applications a considerable subset of interfaces have null type variability. This shows that interfaces are considerably used (overused) to encode shared similarity of classes *belonging to the same type-family*, where abstract classes can be used instead of interfaces. Hence, this shows that Java developers prefer interfaces than abstract to design the abstract types of their concrete implementations. We conjecture that the advantage of interfaces over abstract classes, which is enabling multi-inheritance, lead Java developers to use interfaces even though the multi-inheritance mechanism is not really needed -- at least not in the current version of the software project. Indeed, this point is demonstrated in Figure 1, which shows that most interfaces in studied applications do not have a null type variability. In fact, the largest subset of interfaces are those which have tiny or small type variability. In all these cases, abstract classes cannot be used instead of interfaces. Moreover, the figure shows that interfaces which have a medium to huge type variability are noticeably present in studied applications (specially in large projects such as Argouml and Hibernate). Such interfaces represent shared similarities between (and/or define contracts for) a large variety of classes that belong to different type-families.

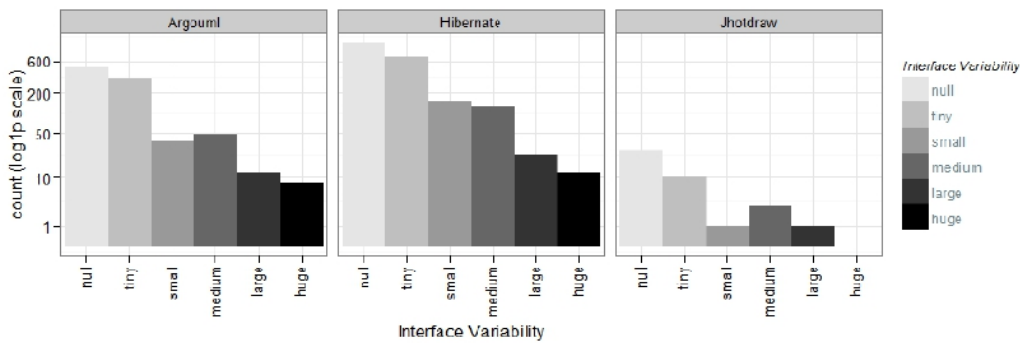

Figure 1. Histogram of Interface Type Variability in studied applications

- *Interface Type Completeness:* Figure 1 shows the histogram of interface type completeness in analyzed applications: (partial, TC < 0.40) (semi-partial, 0.40 ≤ TC ≤ 0.60); (semi-complete, 0.60 < TC < 1); (complete, TC = 1). Surprisingly, the figure shows that for all analyzed applications the presence of interfaces which can be considered as complete types of all their implementations is considerable. More specifically, the figure shows that interfaces which are categorized as complete (or semi-complete) types of their implementations are present in all studied applications as (or more than) those which are partial or semi-partial types. This shows that there is (approximately) a 50% of chance that an interface can be used as a reference type instead of all its implementations, *and this is in all their usage contexts*. Still that for around 25% of interfaces, they are really partial types of their implementations since they specify (focus on) a small subset (a specific functionality) of their implementing classes.

Figure 3 shows that the density of interfaces which have very high TC values and of those which have very low TV values is considerably larger than the density of smaller/greater values of TC/TV. This surprising results show that interfaces are often used to declare all (or almost all) the APIs of their implementing classes rather than focusing on specific functionality. Moreover, they are often used to encode shared similarity between classes *belonging to the same type-family*, rather than being implemented by classes of different type-families. This leads us to conjecture that the motivation beyond using interfaces is not limited to the fact that interfaces enable multi-inheritance mechanism.

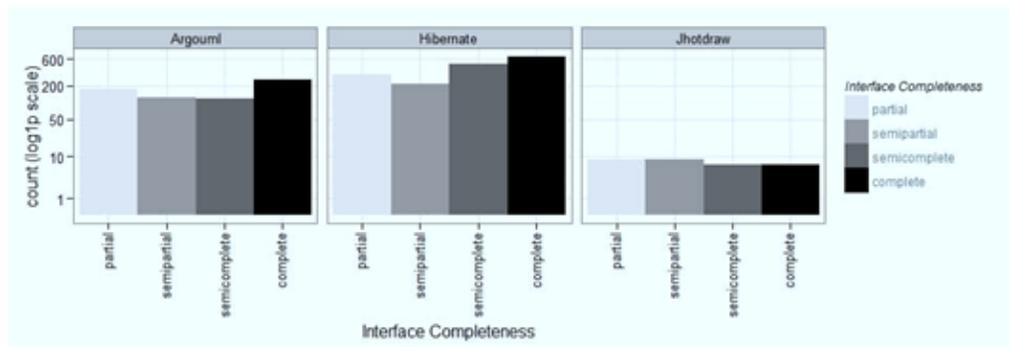

Figure 2. Histogram of Interface Type Completeness in studied applications

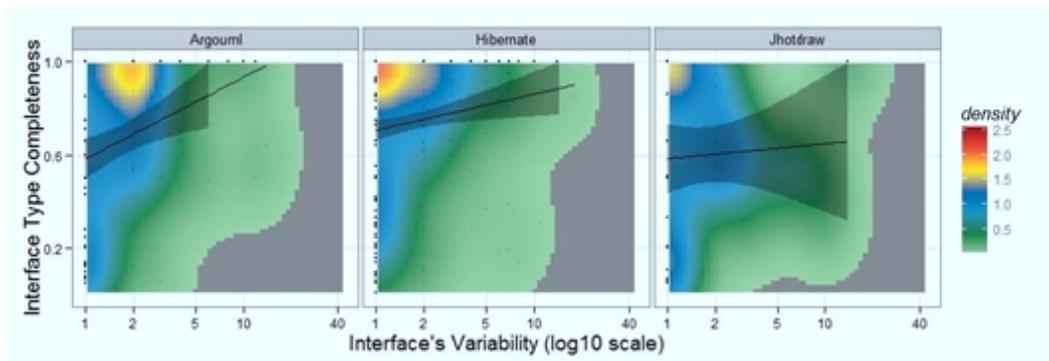

Figure 3. Density 2D of Interface Type Completeness values and Interface Type Variability values for all interfaces in studied applications

## 3. RELATED WORK

The authors in [5] define a set of primitive metrics that measure the complexity and usage of interfaces: the number of interface methods, all arguments of the interface methods, and the number of interface client classes. More complex metrics are defined in [6, 9] to assess existing similarities among interfaces, and evaluate the redundancy in interface sub-class hierarchies.

From another perspective, the authors in [7] investigated the relations between interface duplications and code clones in implementation methods. The study reports that there is a positive correlation between these variables. In [8], the authors investigate the adherence of interfaces' design in real-world software systems to the well known design principles, "Interface Segregation" principle (ISP), "Program to an Interface, not an Implementation" principle (PTIP), and the Cohesion principle. The study reports a strong evidence that software developers definitely abide by the ISP and PTIP, but they neglect the Cohesion property at interface design. It also reports that the negligence of the Cohesion property at interface design leads to a degraded Cohesion in classes which implement interfaces.

## 4. CONCLUSION

In this paper, we empirically analysed the type completeness and variability of interfaces in 3 well known Java projects. For this purpose, we designed two new metrics, the Type Variability (TV) metric and the Type Completeness (TC) metric. Surprisingly, the results showed that interfaces are most often characterized by (very) high type completeness and (very) low type variability.

### ACKNOWLEDGEMENTS

This publication was made possible by NPRP grant #09-1205-2-470 from the Qatar National Research Fund (a member of Qatar Foundation).